\documentclass[10pt]{article}

\usepackage[letterpaper,margin=2cm, columnsep=.5cm]{geometry}
\usepackage{graphicx}
\usepackage{amsmath}
\usepackage{amssymb}
\usepackage{authblk}
\usepackage{pdfpages}

\usepackage[english]{babel}
\usepackage[round, numbers, sort&compress]{natbib}
\usepackage{bibentry}
\nobibliography*
\makeatletter
\renewcommand{\@biblabel}[1]{(#1)}
\makeatother

\makeatletter
\renewcommand{\fnum@figure}{\textbf{Figure \thefigure}}
\renewcommand{\fnum@table}{\textbf{Table \thetable}}
\makeatother

\usepackage{caption}
\usepackage{bm}
\usepackage{stfloats}
\usepackage{color}
\usepackage[title]{appendix} 

\usepackage{setspace}

\usepackage{caption}
\title{\vspace{0mm} 
Evidence on Slowing Progress in Longevity -- Is it Misleading?
}
\author[1]{Silvio C. Patricio}
\author[1]{Annette Baudisch}
\affil[1]{\small Interdisciplinary Center on Population Dynamics, University of Southern Denmark, Campusvej 55, 5230 Odense M, Denmark}
\date{}

\begin{document}
\twocolumn[
\maketitle

\begin{abstract} 
A growing literature argues that progress in human longevity is slowing, and that this slowdown may signal an approaching biological limit to lifespan. The argument rests on visible flattening of cumulative life expectancy records, evaluated through linear regressions or through comparisons of average gains across decades. These summaries, however, provide only weak evidence for slowing improvements, because they cannot distinguish a true from an apparent slowdown. An apparent slowdown may arise due to periods of low variance around unchanged mean annual gains. Applying a Bayesian change-point framework to annual gains, both at the best-practice frontier and separately for each record-holding country, we find no evidence that the expected annual gain has changed across nearly two centuries, but strong evidence that its variance has moved through distinct historical regimes, with progress most regular in recent decades. The recent flattening is consistent with a low-variance regime and provides little evidence for a decline in the underlying mean pace of improvement. Claims that the best-practice frontier signals a biological ceiling should therefore not rest on the recent bend in the cumulative record alone. The frontier is better read as an empirical benchmark, the highest country-specific survival achieved under observed historical conditions.
\end{abstract}

\vspace{2mm}
\noindent \textbf{Keywords:} Best-practice life expectancy $|$ Longevity progress $|$ Population aging $|$ Change-point analysis
\vspace{6mm}

]

\section*{Introduction}

Few trends in modern history have been as steady as the rise in human life expectancy. Two centuries of falling mortality have brought every successive generation a longer life than the preceding one \citep{oeppen2002broken, vaupel2021demographic}. In recent decades, however, the line appears to have started bending. Is this the end of two centuries of sustained gains in human longevity? A growing literature suggests that it might be and that we may be approaching biological limits to lifespan. Reports of stagnating mortality improvements, temporary reversals, and growing uncertainty about future gains have accumulated in high-income populations \citep{ho2018recent, leon2019trends, raleigh2019trends, harper2021declining, dowd2025progress, callaway2025ageing, olshansky2024implausibility, andrade2025cohort}.

The narrative now reaches far beyond academia. Policy reports and popular science writing describe global longevity gains as decelerating, and some frame further extension as constrained by biological aging. Forecasts of pension systems, biomedical research priorities, and individual expectations about future life now rest on the assumption that this slowdown is real \citep{vaupel2021demographic, crimmins2025life, goldman2013substantial, bravo2021addressing}. If recent flattening is treated as a true decline in the expected annual gains, long-run population projections risk embedding lower gains than the actual historical frontier process supports. To properly evaluate whether progress is slowing, the key question is not simply whether the trend is bending but what kind of change generated the bend. 

Current evidence does not address this distinction. The recent literature evaluates longevity trends either by fitting linear or piecewise regressions to the cumulative life expectancy series \citep{oeppen2002broken, vallin2009segmented, vaupel2021demographic, shi2025trend, andrade2025cohort}, or by comparing average annual gains over decades and other fixed windows \citep{cardona2018slowing, olshansky2024implausibility, steel2025changing, dowd2025progress}. When the slope or window average drops, this is taken as evidence of slowing progress. However, both approaches summarize each historical period with a single number. This only allows detecting a bent in the cumulative record, but not its underlying generative mechanism.

A cumulative life expectancy record can flatten for two distinct reasons. The underlying gain process may have slowed on average. Or its yearly gains may have become more regular while their average stays unchanged. Both scenarios differ in mechanism, interpretation, and what they imply for the future of longevity. The former would signal a true decline in the expected pace of progress. The latter instead would signal a change in the regularity of progress, but not necessarily in its pace. This distinction is crucial in concluding whether progress is slowing or not.

\begin{figure}[!htb]
    \centering
    \includegraphics[width=\linewidth]{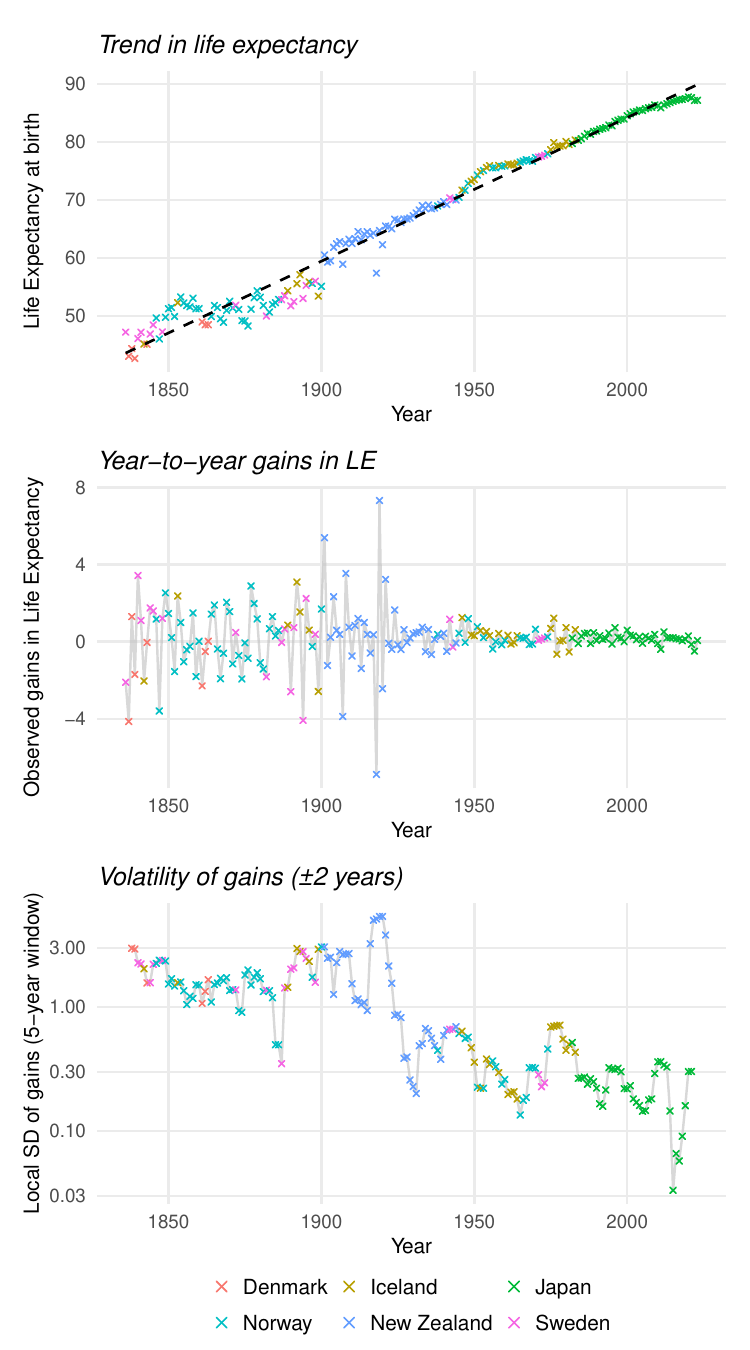}
    \caption{
        \textbf{Best-practice life expectancy at birth, gains, and volatility.}
        Top: Record holders' life expectancy at birth since the mid-nineteenth century. The dashed line shows the linear trend for the best-practice ($\approx$0.25 years/year).
        Middle: Year-to-year gains in life expectancy. Black line shows the gains for the best-practice. Gains fluctuate around a stable mean but show substantial short-term variability.
        Bottom: Local volatility of gains (5-year rolling SD). Black line shows the volatility for the best-practice. Volatility is not constant, showing distinct historical regimes.
    }
    \label{fig:fig_intro}
\end{figure}

As a widely cited example, take the best-practice line of life expectancy at birth, defined by the highest national life expectancy observed each year (Figure \ref{fig:fig_intro}, top panel). This record has risen by roughly 0.25 years annually for nearly two centuries \citep{oeppen2002broken, vaupel2021demographic}. Yet the rise is uneven. The series includes alternating periods of acceleration, deceleration, and stagnation. Japan has dominated the frontier since the late 1980s, and since early 2010s the level series appears to flatten. This flattening is the visual evidence most often cited in support of a longevity slowdown. But does it truly signal a slowing pace of the advancing frontier?

To reveal the generative mechanisms of the frontier, Figure \ref{fig:fig_intro} separates the year-to-year gains of the record (middle panel) from their volatility (bottom panel), the two components that together produce the cumulative trend in the top panel. The gain series measures the frontier directly. Its mean is the expected annual gain, and if that mean has declined, then progress has slowed. The volatility series describes how regular that progress is. High volatility, as in the first century of the record, corresponds to an uneven frontier in which years of large improvement alternate with periods of reversal or stagnation. Low volatility, as in recent decades, corresponds to gains concentrated near their expected value, with progress that is more regular and incremental.

A flatter level series can arise from either a lower expected gain or a lower-volatility regime in which gains stay tightly concentrated around an unchanged expected value. Separating these possibilities requires a model that can detect shifts in the mean and the volatility of the gain process independently.

We use a Bayesian multipartition framework \citep{pedroso2023multipartition} that lets both parameters change at unknown points in time, with the number and location of those points inferred from the data. Our approach builds on a recent move from levels to gains in the analysis of life expectancy \citep{goldstein2024life, bonnet2026potential}. We apply this framework to the composite best-practice gain series and separately to each record-holding country.

\section*{Results}

We analyze annual gains in female best-practice life expectancy at birth using life tables from the Human Mortality Database \citep{hmd}. Because country coverage is sparse in the earliest decades, we restrict the analysis to the period from 1836 onward, keeping only years in which data are available for at least three countries.

We model these gains within a Bayesian multipartition framework \citep{pedroso2023multipartition} in which the mean ($\mu$) and variance ($\sigma^2$) of the process can change independently over time. The number of change points in each parameter, $N_1$ and $N_2$, is treated as unknown and inferred from the data. The posteriors of $N_1$ and $N_2$ therefore quantify, separately, the evidence for shifts in the expected annual gain and in its volatility. Model details and residual diagnostics are reported in the SI Appendix. The diagnostics show no remaining patterns in the residuals over time and no remaining correlation or volatility clustering.

\subsubsection*{The mean of annual gains is stable}

\begin{figure}[!htb]
    \centering
    \includegraphics[width=\linewidth]{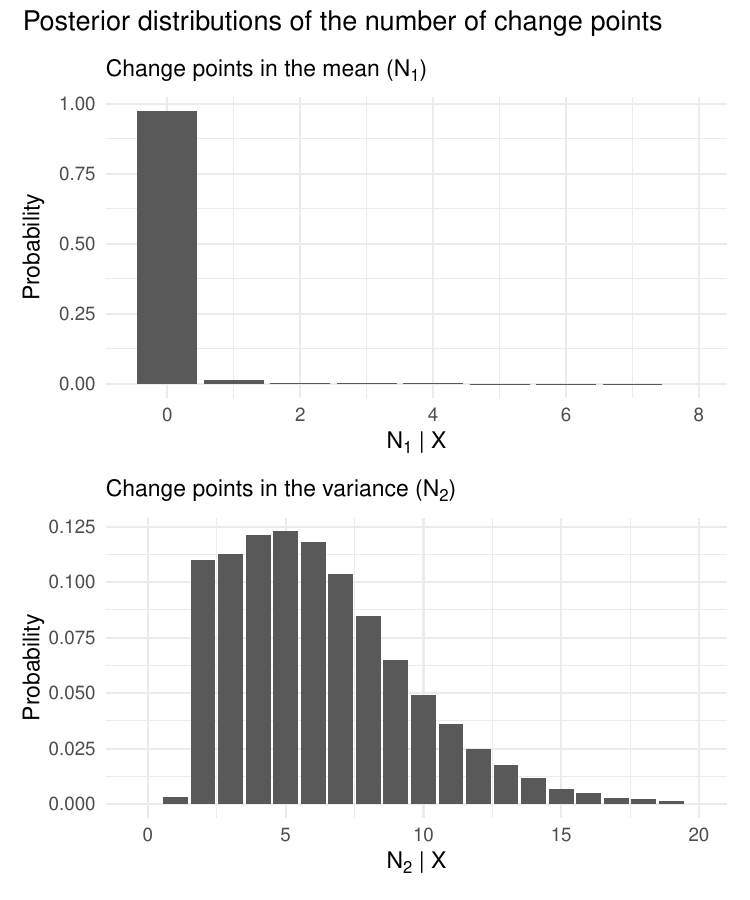}
    \caption{\textbf{Posterior distribution of change points for best-practice life expectancy gains.} Top: The posterior distribution for the number of change points in the mean ($N_1$) is concentrated almost entirely at zero. Bottom: The posterior distribution for the number of change points in the variance ($N_2$) is clearly away from zero, indicating multiple historical regimes of volatility.}
    \label{fig:change_points_distribution}
\end{figure}

The posterior distribution of $N_1$ is concentrated almost entirely at zero (Figure~\ref{fig:change_points_distribution}, top), with an average around 0.0735. Over nearly two centuries, the data provide no support for a change in the expected annual gain. Hence, the frontier is best described as a process whose gains fluctuate around a single common mean, rather than one that has been faster or slower in different historical phases.
 
The top panel of Figure~\ref{fig:results} places the estimated mean, $\mu$, close to $0.1839$ years per year. This sits below the familiar slope ($\beta$) of about $0.25$ years per year fitted to the level series \citep{oeppen2002broken, vaupel2021demographic}, but the two quantities are not directly comparable: $\beta$ depends on how deviations from the mean accumulate and persist over time in the cumulative record, whereas $\mu$ is the expected value of the gain process itself (SI Appendix, Section \textit{Why the slope of the cumulative frontier can differ from the mean gain}). The gap between $\beta$ and $\mu$ reflects how past deviations have accumulated in the level series, not a disagreement between the two estimates. It also explains why segmented fits to the level series \citep{vallin2009segmented} can identify apparent historical phases while the gain process shows no comparable break.

\begin{table}[!htb]
\centering
\caption{Posterior probability of at least one change in the process parameters across different time windows.}
\label{tab:prob_change}
\begin{tabular}{l c c}
\hline
Time Window & $P(N_1 \ge 1 \mid X)$ & $P(N_2 \ge 1 \mid X)$ \\
\hline
Full Series     & 0.0735 & 0.9988 \\
Last 100 Years  & 0.0357 & 0.9936 \\
Last 50 Years   & 0.0211 & 0.9355 \\
Last 25 Years   & 0.0158 & 0.6319 \\
Last 10 Years   & 0.0092 & 0.4070 \\
\hline
\end{tabular}
\end{table}
 
Across all time windows we consider (Table~\ref{tab:prob_change}), we find no evidence of a change in the mean. The posterior probability of at least one change in $\mu$ stays close to zero whether we look at the full series, the last 100, 50, 25, or 10 years. Even in the period in which the cumulative record appears to flatten, the model finds no evidence that the expected annual gain has shifted. This apparent stability holds despite a complete change in the demographic engine of progress over this period, with the main source of gains shifting from improvements in infant and child mortality to adult and older-age mortality \citep{wilmoth2000demography, bongaarts2006long, vallin2004convergences, vaupel2021demographic}. These mechanisms have varied in age and cause, but we find no evidence that their net effect on the expected annual gain has changed.

\subsubsection*{Volatility moves through distinct historical regimes}

\begin{figure}[!htb]
    \centering
    \includegraphics[width=\linewidth]{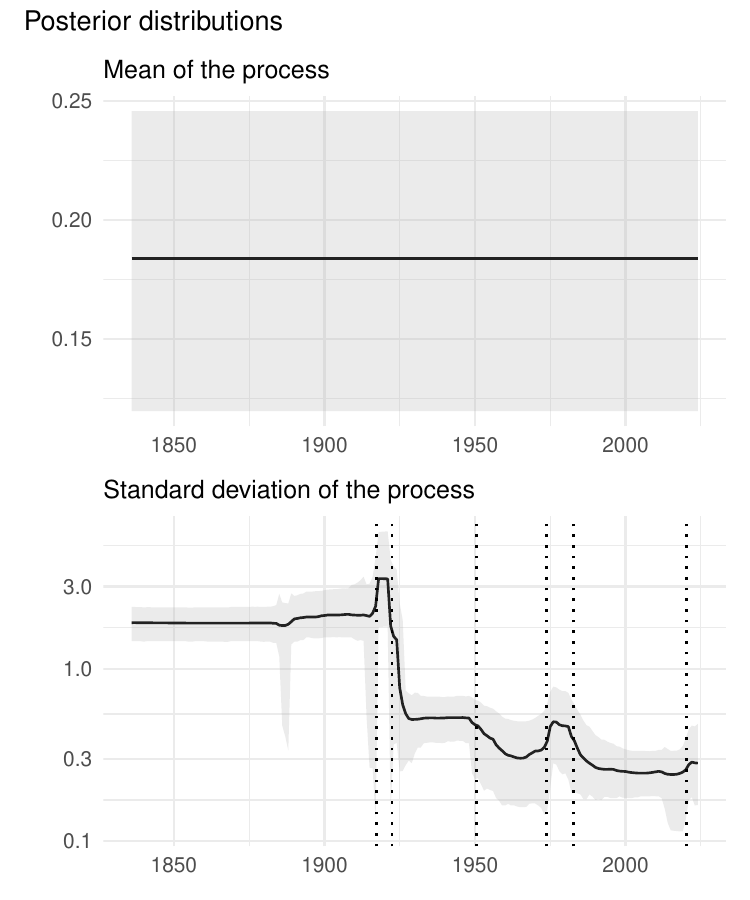}
    \caption{\textbf{Posterior summaries of the stochastic process for annual gains.} Top: The mean of the process remains nearly constant over time. Bottom: The standard deviation shows pronounced historical shifts. Shaded areas represent 95\% highest posterior density intervals. Vertical dotted lines indicate estimated breakpoints in volatility (1917, 1922, 1950, 1974, 1983, and 2020).
    }
    \label{fig:results}
\end{figure}

Unlike the mean, the variance shows clear evidence of change over time. The posterior of $N_2$ is concentrated away from zero, with an average around 6 (Figure~\ref{fig:change_points_distribution}, bottom). And its estimated standard deviation (Figure~\ref{fig:results}, bottom) moves through a clear sequence of historical regimes.

Volatility was high through the nineteenth century, spiked around the First World War and the 1918 influenza pandemic, then dropped in the early 1920s and continued to decline through most of the twentieth century. By the most recent decades, the standard deviation of annual gains is at its lowest level in the entire series.

As gains accumulate, a run of unusually favorable or unfavorable years pulls the cumulative trajectory away from any straight-line summary when volatility is high. When variance is low, the trajectory grows steadily at a pace close to $\mu$. But because $\mu$ is not the same as the slope of a regression line fitted to the entire level series, a low-volatility regime can look like a flattening relative to the historically fitted line, even when the expected annual gain itself has not changed \citep{goldstein2024life}.

Volatility also depends on how often the lead shifts among countries, since the frontier is defined as a maximum. When the frontier passes among many countries whose gains arrive at different times and in different sizes, gains at the top are uneven; when a single country or a small cluster holds the lead for sustained periods, gains become more regular. The first decades of the series are dominated by frequent turnover among Nordic countries and New Zealand; the recent period is dominated by Japan and a small group of other high-longevity populations \citep{vaupel2021demographic}.

\subsubsection*{The frontier mean is not an artifact of composition}
 
A natural concern is that the stability of $\mu$ could come from turnover among countries with different mean gains, rather than from a shared long-run gain among leading populations. To test this, we fit the same multipartition model separately to each country that has held the record for a substantial period (SI Appendix, Figures \ref{fig:denmark_female_0}--\ref{fig:sweden_female_0}). For five of the six record-holding countries (Sweden, Norway, Denmark, Iceland, and New Zealand), the posterior of $N_1$ is again concentrated at zero, with annual gains centered close to the frontier mean even when their levels no longer define the record.

The only exception is Japan, where the model identifies two breakpoints in the mean, near the mid-1950s and the late 1980s. The latter coincides with Japan’s rise to the frontier. Before that shift, Japan’s expected annual gain was approximately $0.37$ years per year, the rate at which it was climbing toward the leading populations. After reaching the lead, its mean dropped to about $0.1839$ years per year, matching the long-run frontier mean.
 
Japan’s slowdown, therefore, reflects its transition from catching up to leading, not a slowdown of the frontier itself. Thus, the absence of a meta-level break in $\mu$ is not an artifact of pooling, but the natural consequence of an accelerating country that upon arrival at the frontier settled at the long-run frontier rate.

\section*{Discussion}

Over nearly two centuries, the data show no evidence supporting a change in the expected annual gain at the best-practice frontier. What has changed is the regularity with which the frontier has advanced. The record has moved through historical regimes of higher and lower volatility, with the lowest volatility seen in recent decades. Apparent flattening of the cumulative record, including the deviation from the long-run trend visible since the early 2000s, is consistent with this low-volatility regime, without any reduction in the expected pace of progress.

 
This finding does not claim that no country has experienced a slowdown, that the demographic engines of progress have stayed the same, or that future gains are guaranteed. It claims something narrower: the historical record of the best-practice frontier carries strong evidence for changing volatility but close to none for shifts in the expected annual gain.

\subsubsection*{Reinterpreting the slowdown debate}

Findings on slowdowns and reversals in high-income populations \citep{ho2018recent, leon2019trends, raleigh2019trends, callaway2025ageing} raise the same question we ask of the frontier itself. Much of that literature rests on either a slope fitted to the cumulative life expectancy record \citep{vallin2009segmented, shi2025trend, andrade2025cohort} or an average growth rate over fixed windows \citep{cardona2018slowing, olshansky2024implausibility, steel2025changing, dowd2025progress}, and neither approach separates the mean of the gain process from its variance. We test this directly for the six record-holding countries, fitting the same model to each (SI Appendix, Figures \ref{fig:denmark_female_0}--\ref{fig:sweden_female_0}).

Five of the six record-holding countries (Denmark, Norway, Sweden, Iceland, and New Zealand) show no change in the mean gain. Their reported slowdowns, including the declines widely documented around 2014 to 2015 \citep{ho2018recent}, are therefore more consistent with a shift in variance than with a shift in pace. Japan is the exception. Even there, its mean gain settled back at the long-run frontier mean rather than falling below it, consistent with a country moving from catching up to leading, not with the frontier itself losing pace.

A stable frontier and a slowing national trajectory are not contradictory in any case, because they describe different objects. A country's own annual gains can slow while the frontier mean stays unchanged. A country can also keep improving every year and still fall behind the frontier if other populations improve faster. Falling behind is therefore relative. A population may gain life expectancy in absolute terms while losing distance to survival levels already reached elsewhere \citep{bonnet2026potential}. Among the record holders in our sample, this is exactly what happens once a country loses the lead but keeps advancing near the long-run frontier mean.

The same caution applies to claims about limits to longevity \citep{olshansky2024implausibility, andrade2025cohort}. Our results do not show that gains will continue indefinitely, and they do not rule out future constraints. They show that recent flattening reflects a change in the regularity of frontier gains, not in their pace, and a narrower set of countries still able to keep up with it.

When the mean is stable, volatility becomes more important for determining who holds the lead at any given moment. A country with high volatility can briefly overtake others through an unusually large gain. A country whose gains stay near a stable mean can hold the lead for longer.
\subsubsection*{Why mean stability is plausible despite changing engines of progress}
 
The stability of $\mu$ holds across nearly two centuries during which the underlying drivers of life expectancy gains have changed. In the nineteenth and early twentieth centuries, progress came mainly from reductions in infant and child mortality \citep{shaw2020introduction, wilmoth2000demography}. Death rates at those ages are highly sensitive to infectious disease cycles and other large shocks, so gains were shaped by a small number of high-impact events with substantial year-to-year volatility.
 
As the main source of progress shifted toward adult and older-age mortality \citep{wilmoth2000demography, bongaarts2006long, vallin2004convergences, vaupel2021demographic, tsui2026analyzing}, the structure of annual gains changed. Gains at older ages reflect the accumulation of many improvements across birth cohorts, causes of death, and ages \citep{mathers2015causes, tsui2026analyzing}. The annual gain at the frontier therefore moved from being shaped by a few large events to being shaped by the sum of many smaller ones.

This shift offers a coherent account of both results together. A more diversified process can produce a more regular annual gain, even when the average value of that gain stays the same. Stability of the mean and falling volatility are therefore compatible with the shift from early-life to later-life mortality improvement \citep{tsui2026analyzing}.

Why these different contributions should net to a stable rate over nearly two centuries remains open. One possible mechanism, likely not the only one, is that progress of any kind, medical, behavioral, or institutional, reaches a leading population only as fast as it can be absorbed.

Adoption costs, infrastructure, and institutional capacity have this effect for medical technologies, clinical practices, and public health policy \citep{hum2015global, flessa2021innovations, adlung2025existing}, and a similar constraint could apply elsewhere. Under that view, the average annual gain at the frontier reflects not the size of any one source, but the speed at which the leading population can absorb and translate it into longer lives. Whether such a constraint produces the observed regularity, or whether the regularity arises from another source, remains a question for future work.



\subsubsection*{Implications for forecasting and interpretation}

The variance of annual gains at the frontier deserves the same forecasting attention as the mean. A model that interprets a low-volatility period as a permanent slowdown will forecast too pessimistically. By the same logic, a model that treats an unusual run of large gains as a new mean regime will forecast too optimistically \citep{goldstein2024life, vaupel2021demographic}.

For forecasters, the relevant object is not only the slope of the cumulative record, but the stochastic process that generates annual gains. For projections of the frontier itself, the long-run mean gain ($\mu \approx 0.1839$ years per year) provides the natural anchor, with recent low-volatility periods treated as changes in variance rather than as new mean regimes. For projections of individual populations, the frontier informs the long-run pace achieved by leading populations, but it does not specify what any single country should be expected to sustain. Forecasting approaches, especially frontier-anchored models \citep{torri2012forecasting, hilton2021modelling, medford2017best}, should therefore separate the expected gain from the regularity of that gain, and test whether apparent bends in the level series reflect a change in the mean, a change in variance, or both.

The stakes accumulate over time. Pension system planning, healthcare expenditure forecasting, and biomedical research prioritization all rest on assumptions about the future trajectory of longevity \citep{vaupel2021demographic, crimmins2025life, goldman2013substantial, bravo2021addressing}, and errors in those assumptions compound over decades. The frontier remains a useful benchmark for comparison, but it is not a direct forecasting rule for any single population.
 
\subsubsection*{What the frontier can and cannot say about biological limits}
 
The frontier record does not directly test the biology of aging. It does, however, weaken one common inference: recent flattening of the best-practice level series is not strong evidence that human longevity gains are approaching a biological ceiling.
 
Under this interpretation, the question is not only whether humanity is approaching a biological limit, but why some populations fail to match survival levels already achieved elsewhere. Recent work on high-income countries points to stalling cardiovascular improvements, external causes of death, widening inequalities, and adverse risk-factor trends as contributors to national slowdowns \citep{harper2021declining, raleigh2019trends, leon2019trends, dowd2025progress, steel2025changing}. Sub-national evidence also shows that vanguard regions can continue to improve at a stable pace while lagging regions slow down \citep{bonnet2026potential}.

These findings support the view that the frontier is an achieved benchmark rather than a biological ceiling. Distance from the frontier may include preventable mortality, because it reflects deaths occurring at ages where survival has already been extended elsewhere. But it may also reflect structural inequalities, cohort histories, measurement differences, and random shocks. Gaps from the frontier therefore mix preventable and unpreventable components, and the frontier is best understood as a record of achieved survival rather than a clean measure of what could be prevented. 


An additional biological hypothesis follows from this view, but it remains interpretive. \citet{patricio2026rhythm} found no sustained directional change in the individual rate of aging across 12 cohort populations, once historical period shocks were accounted for. If the rate of aging is biologically conserved, then longevity gains at the frontier would have to come from postponing mortality rather than slowing senescence. This is consistent with the postponement hypothesis articulated by \citet{vaupel2010biodemography}, where postponement includes both biological delay of onset and reductions in deaths from preventable causes.

Our analysis does not separate these mechanisms. The stable mean gain at the frontier is consistent with a steady joint effect of postponement and death prevention, but it does not prove that such a mechanism produced the pattern. Cohort-level analyses and extensions to a broader set of low-mortality populations are needed to test whether constant aging and stable frontier gains reflect the same underlying process or two regularities that coexist.

\subsubsection*{Limitations and extensions}
 
We do not claim that mean shifts cannot occur in the future or in other mortality series. In the historical best-practice record, the evidence points more clearly to changes in variance than to changes in the expected annual gain. Because the frontier draws from different countries over time, the result cannot be used to dismiss evidence of slowdown in particular countries, regions, age groups, or cohorts. The Japan example shows that country-level mean shifts can occur even while the composite mean stays stable.
 
The model also does not identify the sources of the volatility regimes it detects. These may reflect mortality shocks, wars, pandemics, famines, medical advances, turnover among lead countries, or shifts in the set of populations near the frontier.

What annual gains in life expectancy actually reflect is also an open question. Two countries can sit at different levels of life expectancy but advance at similar gains, or share the same level but differ in gain. Whether the answer is primarily biological, environmental, institutional, or related to the diffusion of medical and behavioral interventions remains open.
 
This analysis uses annual gains in period life expectancy at birth. Cohort life expectancy \citep{andrade2025cohort}, life expectancy at older ages, modal age at death \citep{vazquez2025mortality}, and frontier measures defined for narrower age ranges may behave differently. Linking the frontier approach used here to age-specific decompositions and cohort-based forecasts would help clarify whether the stability we find in the mean of annual gains is a feature of the best-practice record alone or a broader regularity in low-mortality populations.

\section*{Conclusion}

Our analysis finds no evidence that the recent flattening of the longevity record marked the end of two centuries of sustained progress. The expected annual gain at the best-practice frontier has remained stable for nearly two centuries. What has changed is the regularity of those gains, with recent decades showing the lowest volatility. The deviation visible since the early 2000s is therefore better understood as part of a low-volatility regime than as evidence of a slowdown in the underlying pace.

Bends in the cumulative record by themselves do not suffice to deduce a decline in expected annual gains. Interpreting such bends as evidence of a biological slowdown mistakes the visible for the underlying. The burden of evidence therefore shifts. Claims of a longevity slowdown should show a decline in the expected annual gain, not only a visible bend in a cumulative life expectancy record.

The distinction matters for cumulative trend records in general. A frontier that flattens because its gains have become more regular looks, on paper, the same as one that flattens because its gains have slowed. Only the underlying gain process can tell the two apart, and for the best-practice frontier it points to regularity, not decline.

More broadly, the frontier is a record of survival already achieved, not a biological ceiling. The gap between any population and that record is not one thing. Part of it is catch-up potential, survival already reached elsewhere. Part of it reflects structural inequality, cohort history, and chance. These are human-made and historical factors, not biological ones. Neither is proof of what biology permits.

A true slowdown at the frontier would require the expected annual gain itself to fall. Two centuries of data have not shown that. Until they do, the recent bend in the record marks a steadier frontier, not a stalling one.

\section*{Acknowledgments}
\paragraph*{Funding:}
This article is funded by the European Union (ERC, Born Once – Die Once, Grant agreement ID 101043983). Views and opinions expressed are however those of the author(s) only and do not necessarily reflect those of the European Union or the European Research Council Executive Agency. Neither the European Union nor the granting authority can be held responsible for them.

\paragraph*{Author contributions:}
Conceptualization: AB, SP; Methodology, Data Analysis and Visualization	SP; Writing: SP, AB; Supervision and Funding Acquisition: AB.

\paragraph*{Competing interests:} There are no competing interests to declare.

{\footnotesize
\bibliography{bibfile}

@article{oeppen2002broken,
  title={Broken limits to life expectancy},
  author={Oeppen, Jim and Vaupel, James W},
  journal={Science},
  volume={296},
  number={5570},
  pages={1029--1031},
  year={2002},
  publisher={American Association for the Advancement of Science}
}

@article{raleigh2019trends,
  title={Trends in life expectancy in EU and other OECD countries: Why are improvements slowing?},
  author={Raleigh, Veena S},
  year={2019},
  journal = {OECD Health Working Papers},
  publisher={OECD Publishing}
}

@article{vaupel2021demographic,
  title={Demographic perspectives on the rise of longevity},
  author={Vaupel, James W and Villavicencio, Francisco and Bergeron-Boucher, Marie-Pier},
  journal={Proceedings of the National Academy of Sciences},
  volume={118},
  number={9},
  pages={e2019536118},
  year={2021},
  publisher={National Academy of Sciences}
}

@article{vallin2004convergences,
  title={Convergences and divergences in mortality: a new approach of health transition},
  author={Vallin, Jacques and Mesl{\'e}, France},
  journal={Demographic research},
  volume={2},
  pages={11--44},
  year={2004}
}

@article{wilmoth2000demography,
  title={Demography of longevity: past, present, and future trends},
  author={Wilmoth, John R},
  journal={Experimental gerontology},
  volume={35},
  number={9-10},
  pages={1111--1129},
  year={2000},
  publisher={Elsevier}
}

@MISC{hmd,
author = {HMD},
title = {The Human Mortality Database},
howpublished = {http://www.mortality.org/},
year = {2026},
}

@article{olshansky2024implausibility,
  title={Implausibility of radical life extension in humans in the twenty-first century},
  author={Olshansky, S Jay and Willcox, Bradley J and Demetrius, Lloyd and Beltr{\'a}n-S{\'a}nchez, Hiram},
  journal={Nature Aging},
  volume={4},
  number={11},
  pages={1635--1642},
  year={2024},
  publisher={Nature Publishing Group US New York}
}

@article{andrade2025cohort,
  title={Cohort mortality forecasts indicate signs of deceleration in life expectancy gains},
  author={Andrade, Jos{\'e} and Camarda, Carlo Giovanni and Pifarr{\'e} i Arolas, H{\'e}ctor},
  journal={Proceedings of the National Academy of Sciences},
  volume={122},
  number={35},
  pages={e2519179122},
  year={2025},
  publisher={National Academy of Sciences}
}

@article{vallin2009segmented,
  title={The segmented trend line of highest life expectancies},
  author={Vallin, Jacques and Mesl{\'e}, France},
  journal={Population and Development Review},
  volume={35},
  number={1},
  pages={159--187},
  year={2009},
  publisher={Wiley Online Library}
}

@article{cardona2018slowing,
  title={The slowing pace of life expectancy gains since 1950},
  author={Cardona, Carolina and Bishai, David},
  journal={BMC public health},
  volume={18},
  number={1},
  pages={151},
  year={2018},
  publisher={Springer}
}

@article{ho2018recent,
  title={Recent trends in life expectancy across high income countries: retrospective observational study},
  author={Ho, Jessica Y and Hendi, Arun S},
  journal={bmj},
  volume={362},
  year={2018},
  publisher={British Medical Journal Publishing Group}
}

@article{pedroso2023multipartition,
  title={Multipartition model for multiple change point identification},
  author={Pedroso, Ricardo C and Loschi, Rosangela H and Quintana, Fernando Andr{\'e}s},
  journal={TEST},
  volume={32},
  number={2},
  pages={759--783},
  year={2023},
  publisher={Springer}
}

@article{goldstein2024life,
  title={Life Expectancy Reversals in Low-Mortality Populations},
  author={Goldstein, Joshua R and Lee, Ronald D},
  journal={Population and Development Review},
  volume={50},
  number={2},
  pages={437--459},
  year={2024},
  publisher={Wiley Online Library}
}

@article{patricio2026rhythm,
  title={The rhythm of aging: Stability and drift in the individual rate of senescence},
  author={Patricio, Silvio C},
  journal={Proceedings of the National Academy of Sciences},
  volume={123},
  number={15},
  pages={e2528146123},
  year={2026},
  publisher={National Academy of Sciences}
}

@article{leon2019trends,
  title={Trends in life expectancy and age-specific mortality in England and Wales, 1970--2016, in comparison with a set of 22 high-income countries: an analysis of vital statistics data},
  author={Leon, David A and Jdanov, Dmitry A and Shkolnikov, Vladimir M},
  journal={The Lancet Public Health},
  volume={4},
  number={11},
  pages={e575--e582},
  year={2019},
  publisher={Elsevier}
}

@article{dowd2025progress,
  title={Progress stalled? The uncertain future of mortality in high-income countries},
  author={Dowd, Jennifer Beam and Polizzi, Antonino and Tilstra, Andrea M},
  journal={Population and Development Review},
  volume={51},
  number={1},
  pages={257--293},
  year={2025},
  publisher={Wiley Online Library}
}

@article{callaway2025ageing,
  title={Ageing populations: new challenges in longevity},
  author={Callaway, Julia and Strozza, Cosmo and Christensen, Kaare and Doblhammer, Gabriele and Rau, Roland and S{\o}gaard, Jes},
  journal={BMC public health},
  volume={25},
  number={1},
  pages={4395},
  year={2025},
  publisher={Springer}
}

@article{bonnet2026potential,
  title={Potential and challenges for sustainable progress in human longevity},
  author={Bonnet, Florian and Alliger, Ina and Camarda, Carlo-Giovanni and Kl{\"u}sener, Sebastian and Mesl{\'e}, France and M{\"u}hlichen, Michael and Thuilliez, Josselin and Grigoriev, Pavel},
  journal={Nature Communications},
  year={2026},
  publisher={Nature Publishing Group UK London}
}

@article{bongaarts2006long,
  title={How long will we live?},
  author={Bongaarts, John},
  journal={Population and development review},
  pages={605--628},
  year={2006},
  publisher={JSTOR}
}

@article{steel2025changing,
  title={Changing life expectancy in European countries 1990--2021: a subanalysis of causes and risk factors from the Global Burden of Disease Study 2021},
  author={Steel, Nicholas and Bauer-Staeb, Clarissa Maria Mercedes and Ford, John A and Abbafati, Cristiana and Abdalla, Mohammed Altigani and Abdelkader, Atef and Abdi, Parsa and Zu{\~n}iga, Roberto Ariel Abelda{\~n}o and Abiodun, Olugbenga Olusola and Abolhassani, Hassan and others},
  journal={The Lancet Public Health},
  volume={10},
  number={3},
  pages={e172--e188},
  year={2025},
  publisher={Elsevier}
}

@article{hilton2021modelling,
  title={Modelling frontier mortality using Bayesian generalised additive models},
  author={Hilton, Jason and Dodd, Erengul and Forster, Jonathan J and Smith, Peter WF},
  journal={Journal of Official Statistics},
  volume={37},
  number={3},
  pages={569--589},
  year={2021},
  publisher={SAGE Publications Sage UK: London, England}
}

@article{ljung1978measure,
  title={On a measure of lack of fit in time series models},
  author={Ljung, Greta M and Box, George EP},
  journal={Biometrika},
  volume={65},
  number={2},
  pages={297--303},
  year={1978},
  publisher={Oxford University Press}
}

@article{medford2017best,
 ISSN = {14359871, 23637064},
 URL = {http://www.jstor.org/stable/26332157},
 author = {Anthony Medford},
 journal = {Demographic Research},
 pages = {989--1014},
 publisher = {Max-Planck-Gesellschaft zur Foerderung der Wissenschaften},
 title = {Best-practice life expectancy: An extreme value approach},
 urldate = {2026-08-04},
 volume = {36},
 year = {2017}
}

@article{torri2012forecasting,
  title={Forecasting life expectancy in an international context},
  author={Torri, Tiziana and Vaupel, James W},
  journal={International Journal of Forecasting},
  volume={28},
  number={2},
  pages={519--531},
  year={2012},
  publisher={Elsevier}
}

@article{shi2025trend,
  title={Trend breaks in life expectancy in the United States over 120 years and potential sources of future gains},
  author={Shi, Jiaxin and Fletcher, Jason M},
  journal={Population Studies},
  pages={1--18},
  year={2025},
  publisher={Taylor \& Francis}
}

@article{harper2021declining,
  title={Declining life expectancy in the United States: missing the trees for the forest},
  author={Harper, Sam and Riddell, Corinne A and King, Nicholas B},
  journal={Annual review of public health},
  volume={42},
  number={1},
  pages={381--403},
  year={2021},
  publisher={Annual Reviews}
}

@article{tsui2026analyzing,
  title={Analyzing Age-Specific Contributions to Life Expectancy Gains Across Europe},
  author={Tsui, Iona and Lee, Yeonjung and Yon, Yongjie and Huber, Manfred},
  journal={Journal of Applied Gerontology},
  pages={07334648261416182},
  year={2026},
  publisher={SAGE Publications Sage CA: Los Angeles, CA}
}

@article{vaupel2010biodemography,
  title={Biodemography of human ageing},
  author={Vaupel, James W},
  journal={Nature},
  volume={464},
  number={7288},
  pages={536--542},
  year={2010},
  publisher={Nature Publishing Group UK London}
}

@misc{vazquez2025mortality,
title = "Mortality {\`a} la mode",
author = "Paola V{\'a}zquez-Castillo",
year = "2025",
month = jun,
day = "16",
doi = "10.21996/5632f35b-0a8f-4eed-bec0-140a5d6917fe",
language = "English",
publisher = "Syddansk Universitet. Det Samfundsvidenskabelige Fakultet",
school = "SDU",
}

@article{goldman2013substantial,
  title={Substantial health and economic returns from delayed aging may warrant a new focus for medical research},
  author={Goldman, Dana P and Cutler, David and Rowe, John W and Michaud, Pierre-Carl and Sullivan, Jeffrey and Peneva, Desi and Olshansky, S Jay},
  journal={Health affairs},
  volume={32},
  number={10},
  pages={1698--1705},
  year={2013}
}

@article{crimmins2025life,
  title={Life expectancy and health expectancy in the Twenty-first century: The unthinkable, the inconceivable, and the unknowable},
  author={Crimmins, Eileen M},
  journal={Demography},
  volume={62},
  number={4},
  pages={1217--1236},
  year={2025},
  publisher={Duke University Press}
}

@article{bravo2021addressing,
  title={Addressing the life expectancy gap in pension policy},
  author={Bravo, Jorge M and Ayuso, Mercedes and Holzmann, Robert and Palmer, Edward},
  journal={Insurance: Mathematics and Economics},
  volume={99},
  pages={200--221},
  year={2021},
  publisher={Elsevier}
}

@misc{shaw2020introduction,
  title={An introduction to the history of infectious diseases, epidemics and the early phases of the long-run decline in mortality},
  author={Shaw-Taylor, Leigh},
  journal={The Economic History Review},
  volume={73},
  number={3},
  pages={E1--E19},
  year={2020},
  publisher={Wiley Online Library}
}

@article{mathers2015causes,
  title={Causes of international increases in older age life expectancy},
  author={Mathers, Colin D and Stevens, Gretchen A and Boerma, Ties and White, Richard A and Tobias, Martin I},
  journal={The Lancet},
  volume={385},
  number={9967},
  pages={540--548},
  year={2015},
  publisher={Elsevier}
}

@incollection{durbin1992testing,
  title={Testing for serial correlation in least squares regression. I},
  author={Durbin, James and Watson, Geoffrey S},
  booktitle={Breakthroughs in statistics: Methodology and distribution},
  pages={237--259},
  year={1992},
  publisher={Springer}
}

@article{wald1940test,
  title={On a test whether two samples are from the same population},
  author={Wald, Abraham and Wolfowitz, Jacob},
  journal={The Annals of Mathematical Statistics},
  volume={11},
  number={2},
  pages={147--162},
  year={1940},
  publisher={JSTOR}
}

@article{breusch1978testing,
  title={Testing for autocorrelation in dynamic linear models.},
  author={Breusch, Trevor S},
  journal={Australian economic papers},
  volume={17},
  number={31},
  pages={334},
  year={1978}
}

@article{godfrey1978testing,
  title={Testing against general autoregressive and moving average error models when the regressors include lagged dependent variables},
  author={Godfrey, Leslie G},
  journal={Econometrica: Journal of the Econometric Society},
  pages={1293--1301},
  year={1978},
  publisher={JSTOR}
}

@article{white1980heteroskedasticity,
  title={A heteroskedasticity-consistent covariance matrix estimator and a direct test for heteroskedasticity},
  author={White, Halbert},
  journal={Econometrica: journal of the Econometric Society},
  pages={817--838},
  year={1980},
  publisher={JSTOR}
}

@article{koenker1981note,
  title={A note on studentizing a test for heteroscedasticity},
  author={Koenker, Roger},
  journal={Journal of econometrics},
  volume={17},
  number={1},
  pages={107--112},
  year={1981},
  publisher={Elsevier}
}

@article{breusch1979simple,
  title={A simple test for heteroscedasticity and random coefficient variation},
  author={Breusch, Trevor S and Pagan, Adrian R},
  journal={Econometrica: Journal of the econometric society},
  pages={1287--1294},
  year={1979},
  publisher={JSTOR}
}

@book{tsay2010analysis,
  title={Analysis of financial time series},
  author={Tsay, Ruey S and Tsay, Ruey S},
  volume={3},
  year={2010},
  publisher={wiley New York}
}

@article{adlung2025existing,
  title={Existing and emerging frameworks for the adoption and diffusion of medical devices and equipment in low-resource settings: a scoping review},
  author={Adlung, Christopher and van der Kooij, Nienke and Diehl, Jan Carel and Hinrichs-Krapels, Saba},
  journal={Health and Technology},
  volume={15},
  number={2},
  pages={273--297},
  year={2025},
  publisher={Springer}
}

@article{hum2015global,
  title={Are global and regional improvements in life expectancy and in child, adult and senior survival slowing?},
  author={Hum, Ryan J and Verguet, St{\'e}phane and Cheng, Yu-Ling and McGahan, Anita M and Jha, Prabhat},
  journal={PLoS One},
  volume={10},
  number={5},
  pages={e0124479},
  year={2015},
  publisher={Public Library of Science San Francisco, CA USA}
}

@article{flessa2021innovations,
  title={Innovations in health care—a conceptual framework},
  author={Flessa, Steffen and Huebner, Claudia},
  journal={International journal of environmental research and public health},
  volume={18},
  number={19},
  pages={10026},
  year={2021},
  publisher={MDPI}
}

@misc{cabral_patricio_2026_21833625,
  author       = {Cabral Patricio, Silvio and
                  Baudisch, Annette},
  title        = {Code for: "Evidence on Slowing Progress in
                   Longevity — Is it Misleading?"
                  },
  day          = 07,
  month        = aug,
  year         = 2026,
  publisher    = {Zenodo},
  version      = {V1.0},
  doi          = {10.5281/zenodo.21833625},
  url          = {https://doi.org/10.5281/zenodo.21833625},
}

@article{nelson1984pitfalls,
  title={Pitfalls in the Use of Time as an Explanatory Variable in Regression},
  author={Nelson, Charles R and Kang, Heejoon},
  journal={Journal of Business \& Economic Statistics},
  volume={2},
  number={1},
  pages={73--82},
  year={1984},
  publisher={Taylor \& Francis}
}
}


\cleardoublepage



\renewcommand{\thefigure}{S\arabic{figure}}
\renewcommand{\thetable}{S\arabic{table}}
\renewcommand{\theequation}{S\arabic{equation}}
\renewcommand{\thepage}{S\arabic{page}}
\setcounter{figure}{0}
\setcounter{table}{0}
\setcounter{equation}{0}
\setcounter{page}{1} 
\begin{appendices}

\section*{Supporting Information}

\subsection*{Database}
We use period life tables from the Human Mortality Database \citep{hmd}, drawing on all countries and calendar years with complete data. The countries included in our sample align with those analyzed by \citet{vaupel2021demographic}. 


Our analysis focuses on a subset of countries that meet two key criteria: long historical coverage and consistently high data quality. These include: Australia, Austria, Belgium, Canada, Switzerland, Czech Republic, Denmark, Spain, Finland, France (total population), Germany (unified or total population), Italy, Japan, the Netherlands, Norway, Portugal, Sweden, the United Kingdom (England and Wales), and the United States. These countries offer extended and uninterrupted series of high-quality life tables, making them suitable for tracking long-term trends in the upper bounds of human longevity. Countries with only recent data, incomplete historical coverage, or regional overlaps (such as separate entries for East/West Germany or subpopulations within the UK or New Zealand) are excluded to ensure consistency.


\subsection*{Methodology}
We analyze annual gains in best-practice life expectancy at birth. For each year, best-practice life expectancy denoted $e^*_0$ is defined as the highest observed female $e_0$ among countries in the Human Mortality Database \cite{hmd}. Because country coverage is sparse in the earliest decades, we restrict the analysis to years from 1836 onward with data available for at least three countries, yielding a series of yearly frontier gains $y_t = e_{0,t}^* - e_{0,t-1}^*$.

To study whether the statistical properties of these gains changed over time, we use the Bayesian multipartition change-point model \cite{pedroso2023multipartition} for normally distributed data with unknown mean and variance. The annual gains are modeled as
\begin{equation*}
y_t \mid \mu_t,\sigma_t^2 \sim \mathcal{N}(\mu_t,\sigma_t^2), \qquad t=1,\dots,n,
\end{equation*}
where the mean $\mu_t$ and variance $\sigma_t^2$ are each assumed piecewise constant over time. A standard change-point model would force $\mu_t$ and $\sigma_t^2$ to share the same breakpoints, so a change in the pace of progress and a change in its regularity would be indistinguishable, or a variance-only shift could be mistaken for a mean shift. The multipartition model removes this constraint by giving $\mu_t$ and $\sigma_t^2$ independent partitions, each with its own unknown number and location of change points, so the model can ask separately whether the frontier's pace changed and whether its regularity changed, rather than answering one blended question.

A cluster in this model is a run of consecutive years compatible with a common mean or a common variance. For the mean process, a cluster is a stretch of years sharing the same expected annual gain, the pace at which the frontier advanced. For the variance process, it is a stretch of years sharing the same spread of gains around that pace, how regular or erratic year-to-year progress was. A change point in $\mu_t$ therefore means a genuine shift in the pace of frontier progress, while a change point in $\sigma_t^2$ means a shift in how evenly that progress arrived, with no implication that the pace itself moved. Both kinds of shifts are most plausible where the demographic conditions behind the frontier changed, most directly when the record passed to a different country.

Posterior inference is based on Markov chain Monte Carlo using the partially collapsed Gibbs sampler described by Pedroso et al. \cite{pedroso2023multipartition}. From the posterior output, we summarize the number and timing of change points in the mean and variance, as well as the corresponding posterior distributions of $\mu_t$ and $\sigma_t^2$ over time.

\subsubsection*{Prior specification}
The model requires priors for the regime-specific means and variances, and for the partition processes governing changes in these parameters. We applied the same prior standard across the composite frontier series and each country-level series, so that differences in the results across series reflect the data rather than a change in the prior underneath them.

For the partition priors, we elicited the Beta hyperparameters $(\alpha_k,\beta_k)$ through their implied prior on the number of change points, following \cite{pedroso2023multipartition}. If the transition probability $p_k$ is $\text{Beta}(\alpha_k,\beta_k)$, the induced prior on $N_k$, the number of change points for parameter $k$ ($k=1$ mean, $k=2$ variance), is Beta-Binomial with
\begin{equation*}
E(N_k)=(n-1)\frac{\alpha_k}{\alpha_k+\beta_k}.
\end{equation*}
For each $k$ we fixed a target $E(N_k)=\nu_k$ and solved for $\beta_k=\alpha_k\frac{(n-1)-\nu_k}{\nu_k}$, keeping the target itself, not an abstract hyperparameter pair, under direct control.

The two targets were set asymmetrically, and deliberately so $\nu_1=5$ is anchored to the six countries
that have held the best-practice record, Denmark, Iceland, Japan, New Zealand, Norway, and Sweden, since a genuine shift in the pace of progress is most plausible when the demographic conditions defining the record itself change. $\nu_2=15$ reflects that the record changed hands 50 times over the study period; we set $\nu_2$ well below that count, since not every change of leader marks a new volatility regime, two successive leaders can share similar demographic conditions, while keeping it well above $\nu_1$. This asymmetry matches what is already visible without any model: annual gains fluctuate with no obvious pattern, while their volatility visibly moves through distinct historical episodes.

The concentration parameters $\alpha_k$ set how strongly the prior holds to these targets. Because $\mathrm{Var}(N_k)$ decreases monotonically in $\alpha_k$ for fixed $\nu_k$ \cite{pedroso2023multipartition}, a small $\alpha_k$ keeps the prior centered on $\nu_k$ while remaining diffuse around it. We used $\alpha_1=0.02$ for the mean process, reflecting genuine uncertainty about whether the pace of progress has changed at all, and a larger $\alpha_2=0.28$ for the variance process, since the historical record already shows visible regime structure that a weaker prior would understate.

For the regime-specific means, $\mu_j \sim \mathcal{N}(\mu_0, S_0^2)$, with $\mu_0$ set to the observed series mean and $S_0^2 \approx 0.14$, the empirical variance of the 5-year rolling mean of the gain series, the same window used for the rolling volatility shown in Figure \ref{fig:fig_intro}. This keeps the prior centered on a demographically reasonable pace of improvement while remaining wide enough to let short or unusual regimes speak for themselves, without assigning meaningful probability to values the frontier could not sustain, such as a persistently negative average gain, which a running maximum of national life expectancies cannot produce over an extended period.

For the regime-specific variances, $\sigma_j^2 \sim IG(a,d)$ with $(a,d)=(0.1,2.1)$, the reasonably flat default proposed by Pedroso et al. \cite{pedroso2023multipartition}. This keeps the variance prior weakly informative, appropriate given that volatility is harder to pin down from a short run of annual gains than the mean is.

\subsection*{Code and Data Availability}

The R code used to assemble the best-practice frontier, fit the Bayesian multipartition change-point model, and generate the figures and tables, is available on GitHub and archived on Zenodo \citep{cabral_patricio_2026_21833625}. The underlying life table data are publicly available from the Human Mortality Database \citep{hmd} upon free registration.

\subsection*{Checking assumptions of previous studies}

\begin{figure}[!htb]
    \centering
    \includegraphics[width=\linewidth]{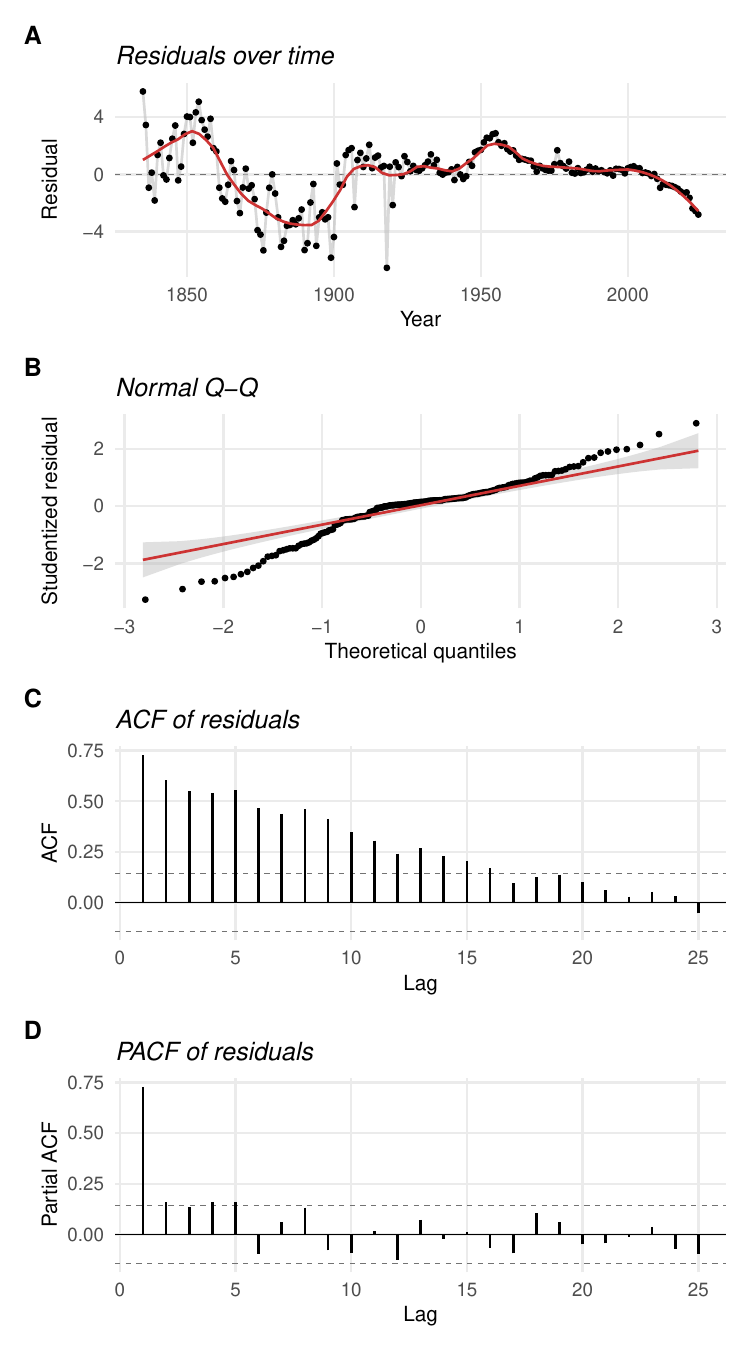}
    \caption{Residual diagnostics for the OLS trend fitted to best-practice life expectancy ($e^{*}_0 \sim \text{Year}$, $n = 190$). (A)~Residuals against calendar year, with a LOESS smoother. (B)~Normal Q--Q plot of the studentized residuals. (C,~D)~Autocorrelation and partial autocorrelation functions of the residuals; dashed lines mark the $95\%$ white-noise bounds.} 
    \label{fig:diagnostics}
\end{figure}

The slowdown evidence in previous studies discussed in the main text is read from an OLS slope fitted to the cumulative best-practice record. We check whether that fit's assumptions hold by regressing $e^{*}_0$ on calendar time ($e^{*}_{0,t} = \alpha + \beta\,t + \varepsilon_t$, $n = 190$); the formal tests are in Table~\ref{tab:diagnostics} and the residual diagnostics in Figure~\ref{fig:diagnostics}. At this sample size the tests reject small departures, so we read them together with the effect sizes. 

\begin{table}[!htb]
    \centering
    \caption{\label{tab:diagnostics} Formal diagnostic tests for the OLS fit $e^{*}_{0,t} = \alpha + \beta\,t + \varepsilon_t$ ($n=190$). At this sample size the tests are powered to detect negligible departures; effect sizes are reported to convey magnitude. $\hat\rho_1$ is the lag-1 residual autocorrelation, $R^2_{\mathrm{aux}}$ the fraction of squared-residual variation explained by Year, and $g_1,g_2$ the residual skewness and excess kurtosis.}
    \centering
        \resizebox{\textwidth/2}{!}{\begin{tabular}[t]{lrrl}
        \hline
            Diagnostic & Statistic & $p$-value & Effect size\\
            \hline
            \multicolumn{4}{l}{\textit{Independence of errors}}\\
            \hspace{1em}Durbin--Watson (lag 1) & 0.492 & $6.7\times10^{-26}$ & $\hat\rho_1 = 0.73$\\
            \hspace{1em}Breusch--Godfrey (lags 1--5) & 110 & $3.4\times10^{-22}$ & \\
            \hspace{1em}Runs test (residual signs) & -6.98 & $2.9\times10^{-12}$ & \\
            \multicolumn{4}{l}{\textit{Conditional variance}}\\
            \hspace{1em}Ljung--Box on $e_t^2$ (10 lags) & 144 & $<10^{-300}$ & \\
            \multicolumn{4}{l}{\textit{Homoskedasticity}}\\
            \hspace{1em}Breusch--Pagan (Koenker) & 30.2 & $4.0\times10^{-8}$ & $R^2_{\mathrm{aux}} = 0.16$\\
            \hspace{1em}White ($\mathrm{Year},\ \mathrm{Year}^2$) & 30.5 & $2.4\times10^{-7}$ & \\
            \hspace{1em}Score test (non-const. var.) & 44.6 & $2.4\times10^{-11}$ & \\
            \multicolumn{4}{l}{\textit{Normality}}\\
            \hspace{1em}Shapiro--Wilk & 0.954 & $8.2\times10^{-6}$ & $g_1 = -0.55,\ g_2 = 0.92$\\
            \hline
            \multicolumn{4}{l}{\footnotesize{\rule{0pt}{1em}\textit{Note:} Sequential tests use residuals ordered by calendar year.}}\\
    \end{tabular}}
\end{table}

\paragraph{Independence of the errors.}
The no-autocorrelation assumption fails. Durbin--Watson is $0.49$ \cite{durbin1992testing}, a lag-one residual correlation of $\hat\rho_1 = 0.73$; Breusch--Godfrey shows the dependence runs past the first lag \cite{breusch1978testing, godfrey1978testing}, and the Runs test ($z = -6.98$) \cite{wald1940test} finds long same-sign stretches, seen as the slow wave in Figure~\ref{fig:diagnostics} panel A. The autocorrelation function ACF falls off gradually and the partial autocorrelation function PACF carries almost all its weight at lag one (Figure~\ref{fig:diagnostics}, panels C,D). This is expected: because $e^{*}_0$ is a running sum of the annual gains, a run of good or bad years accumulates and holds the series off any straight line, so deviations persist. It is the mechanism we discuss in section \emph{Why the slope of the cumulative frontier can differ from the mean gain}. 

\paragraph{Conditional variance.}
The squared residuals are autocorrelated too. A Ljung--Box test on $\varepsilon_t^2$ is strongly significant \cite{ljung1978measure}, evidence of volatility clustering \cite{tsay2010analysis}: large residuals fall in the early record, small ones in recent decades. This is the same volatility regime structure the change-point model assigns to $\sigma^2$.

\paragraph{Heteroskedasticity and the differenced series.}
The error variance is not constant. Breusch--Pagan \cite{breusch1979simple, koenker1981note}, White \cite{white1980heteroskedasticity}, and the score test all reject; the Breusch--Pagan auxiliary regression ties $16\%$ of the squared-residual variation to calendar time, and the spread narrows from the nineteenth century onward (Figure~\ref{fig:diagnostics}, panel A). Differencing the levels returns the annual gains $y_t = e^{*}_{0,t} - e^{*}_{0,t-1}$, whose variance is also uneven and concentrated in the early segments. A model with fixed error variance misses this in both directions.

\paragraph{Normality.}
Shapiro--Wilk rejects normality, but the departure is small (skewness $-0.55$, excess kurtosis $0.92$) and the Q--Q plot is near-linear apart from the tails (Figure~\ref{fig:diagnostics}, panel B). At this sample size the rejection is expected, and it mostly follows from the dependence and changing variance above. 

These failures are structured, not random: the errors are autocorrelated because the level cumulates the gains, and their variance tracks the shifting volatility of those gains. So $\hat\beta$ mixes the mean pace of improvement with the accumulated history of deviations, and cannot be read as the expected annual gain. 

\subsection*{Why the slope of the cumulative frontier can differ from the mean gain}
Let $y_t$ denote the annual gain in best-practice life expectancy at year $t$, and write
\begin{equation*}
y_t = \mu + u_t,
\end{equation*}
where $\mu$ is the mean gain and $u_t$ is the deviation from that mean. The frontier level is the cumulative sum of these gains,
\begin{equation*}
L_t = L_0 + \sum_{s=1}^t y_s = L_0 + t\mu + \sum_{s=1}^t u_s.
\end{equation*}

The second equality splits $L_t$ into a deterministic trend, $t\mu$, and an accumulated noise term, $\sum_{s=1}^t u_s$. Both pieces matter for what follows.
Consider the ordinary least-squares slope obtained by regressing $L_t$ on calendar time $t = 1, \dots, T$. Write $\bar t = \frac{1}{T}\sum_{t=1}^T t = \frac{T+1}{2}$ for the mean of calendar time, and
\begin{equation*}
S_{tt} = \sum_{t=1}^T (t-\bar t)^2
\end{equation*}
for the total variation in calendar time. The slope is then
\begin{equation*}
\hat\beta = \frac{\sum_{t=1}^T (t-\bar t)L_t}{S_{tt}}.
\end{equation*}

To see what $\hat\beta$ estimates, substitute the cumulative representation of $L_t$ into the numerator and expand it term by term,
\begin{equation*}
\sum_{t=1}^T (t-\bar t)L_t
=
  L_0\sum_{t=1}^T (t-\bar t)
+ \mu\sum_{t=1}^T (t-\bar t)\,t
+ \sum_{t=1}^T (t-\bar t)\sum_{s=1}^t u_s.
\end{equation*}

The first term vanishes because centered values sum to zero, $\sum_{t=1}^T (t-\bar t) = 0$, so the starting level $L_0$ drops out and has no bearing on the slope. The second term reduces to $\mu S_{tt}$, because writing $t = (t-\bar t)+\bar t$ shows $\sum_{t=1}^T (t-\bar t)\,t = \sum_{t=1}^T (t-\bar t)^2 + \bar t\sum_{t=1}^T (t-\bar t) = S_{tt}$.
The third term takes more work because it is a double sum. For each year $t$, the inner sum $\sum_{s=1}^t u_s$ collects every deviation up to and including $t$. Equivalently, a given deviation $u_s$ appears in this double sum once for every $t \geq s$, since it enters the cumulative level at time $s$ and stays part of it at every later time. Grouping terms by $s$ instead of $t$, that is, summing over which years each $u_s$ affects rather than which deviations each year contains, gives
\begin{equation*}
\sum_{t=1}^T (t-\bar t)\sum_{s=1}^t u_s
=
  \sum_{s=1}^T u_s\sum_{t=s}^T (t-\bar t).
\end{equation*}
Collecting the three terms and dividing by $S_{tt}$ gives
\begin{equation*}
\hat\beta
=
  \mu+
  \frac{1}{S_{tt}}
\sum_{s=1}^T
\left[
  \sum_{t=s}^T (t-\bar t)
  \right]u_s.
\end{equation*}
This expression already contains the main result. The slope equals the mean gain $\mu$ plus a weighted sum of the deviations $u_s$, where the weight on each deviation is $\sum_{t=s}^T (t-\bar t)$, a decomposition that follows the same logic Nelson and Kang \citep{nelson1984pitfalls} developed for trend-fitted random walks.
That inner sum has a closed form. Since $\sum_{t=s}^T t$ counts the $T-s+1$ integers from $s$ to $T$,
\begin{equation*}
\sum_{t=s}^T t = \frac{T(T+1)}{2}-\frac{(s-1)s}{2},
\end{equation*}
and since $\bar t = \frac{T+1}{2}$, we can write
\begin{eqnarray*}
\sum_{t=s}^T (t-\bar t)
&=&
  \sum_{t=s}^T t - (T-s+1)\bar t \\
&=&
  \frac{(s-1)(T-s+1)}{2}.
\end{eqnarray*}
Substituting this closed form for the weight gives the final result,
\begin{equation*}
\hat\beta
=
  \mu+
  \frac{1}{S_{tt}}
\sum_{s=1}^T
\frac{(s-1)(T-s+1)}{2}\,u_s.
\end{equation*}
This expression shows that the fitted slope of the cumulative level series equals the mean annual gain plus a weighted sum of deviations from that mean. The weight
\begin{equation*}
\frac{(s-1)(T-s+1)}{2}
\end{equation*}
is zero for the very first observation ($s=1$), rises toward the middle of the series, and falls again toward the end. Deviations near the middle of the sample therefore carry the most weight, and deviations near either boundary carry the least. A deviation early in the series still has the whole rest of the trajectory to reassert the mean, and a deviation at the very end has had little time to accumulate, but a deviation near the middle pulls the fitted line off course for longest. As a result, the slope equals $\mu$ only when these weighted deviations cancel out. Runs of positive deviations push the slope above $\mu$, whereas runs of negative deviations push it below. Positive temporal dependence in the gain process does not mechanically create a gap between $\hat\beta$ and $\mu$, but it makes sustained runs more likely and therefore increases the chance of a substantial difference between the two quantities.

The residuals from the OLS fit to the best-practice level series (Figure \ref{fig:diagnostics}, panel A) trace out exactly this kind of run, consistent with the pseudo-cyclical detrending artifact documented by \citet{nelson1984pitfalls}. They are positive through the mid-nineteenth century, persistently negative from the 1870s to 1900, close to zero for most of the twentieth century with a modest positive bump around mid-century, and negative again over the most recent decades. Once the fitted linear trend is removed, these residuals are the running sum $\sum_{s \leq t} u_s$, so their sustained swings above and below zero are the same accumulated deviations that pull $\hat\beta$ away from $\mu$.

In this sense, the slope of the level series and the mean of the gain process are not competing estimates of the same parameter. The former summarizes observed cumulative change, whereas the latter describes the expected annual gain generating that change.

\cleardoublepage
\section*{Supplementary Results and Model Diagnostics}

This section reports supplementary diagnostics for the model presented in the main text, followed by country-by-country results. We first assess the adequacy of the composite frontier model, then present the same diagnostics separately for each of the six record-holding countries.

\subsection*{Model adequacy for the composite frontier}

\begin{figure}[!ht]
    \centering
    \includegraphics[width=\linewidth]{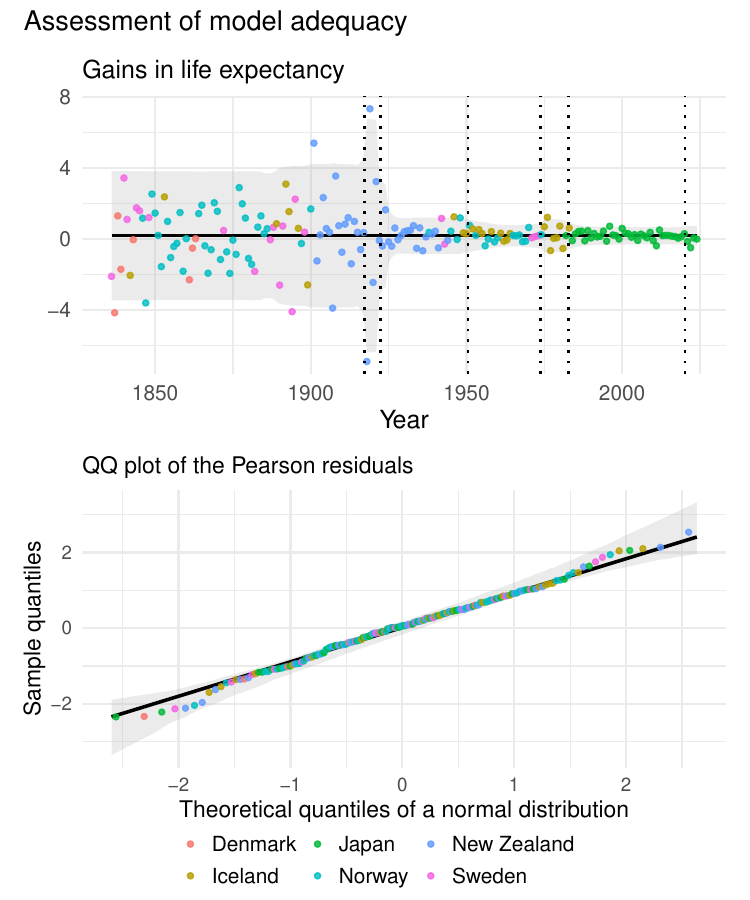}
    \caption{
        \textbf{Observed gains and residual diagnostics.} Top: Observed yearly gains in best-practice life expectancy and fitted distributions by cluster. Dots show the annual gains, colored by the country defining the frontier in each year. The horizontal black line is the estimated mean and the shaded band gives the central 95\% interval. Vertical dotted lines mark estimated breakpoints in volatility. Bottom: QQ plot of the Pearson residuals against a standard normal distribution. The near-linearity of the points supports the distributional assumptions of the model.
    }
    \label{fig:model_adequacy}
\end{figure}

To assess model adequacy, we examine the Pearson residuals
\begin{equation*}
r_t = \frac{y_t - \mu_t}{\sqrt{\sigma_t^2}},
\end{equation*}
which, under the model, should follow a standard normal distribution. The QQ plot in the bottom panel of Figure~\ref{fig:model_adequacy} aligns closely with the theoretical reference over nearly the entire range, with only minor deviations in the extreme tails. Formal diagnostics show no evidence of remaining serial correlation or  autoregressive conditional heteroskedasticity (ARCH) effects (SI Appendix). The dominant temporal structure of the series (the stability of $\mu$ and the regime structure of $\sigma^2$) is therefore well captured by the model.

\subsubsection*{Residual diagnostics}
Residual diagnostics were also examined formally. A Ljung--Bo test applied to the Pearson residuals at lag 10 failed to reject the null of no autocorrelation with $p=0.4351$ \citep{ljung1978measure}. A Breusch--Godfrey test up to order 10 likewise found no evidence of serial correlation with $p=0.3601$ \citep{breusch1978testing, godfrey1978testing}. Finally, an ARCH LM test failed to reject the null of no remaining ARCH effects with $p=0.2118$ \citep{tsay2010analysis}. Together, these results suggest that the fitted model leaves little evidence of residual linear dependence or volatility clustering.

\subsection*{Country-level results}
\begin{figure*}[htbp]
    \centering
    \includegraphics[width=\textwidth]{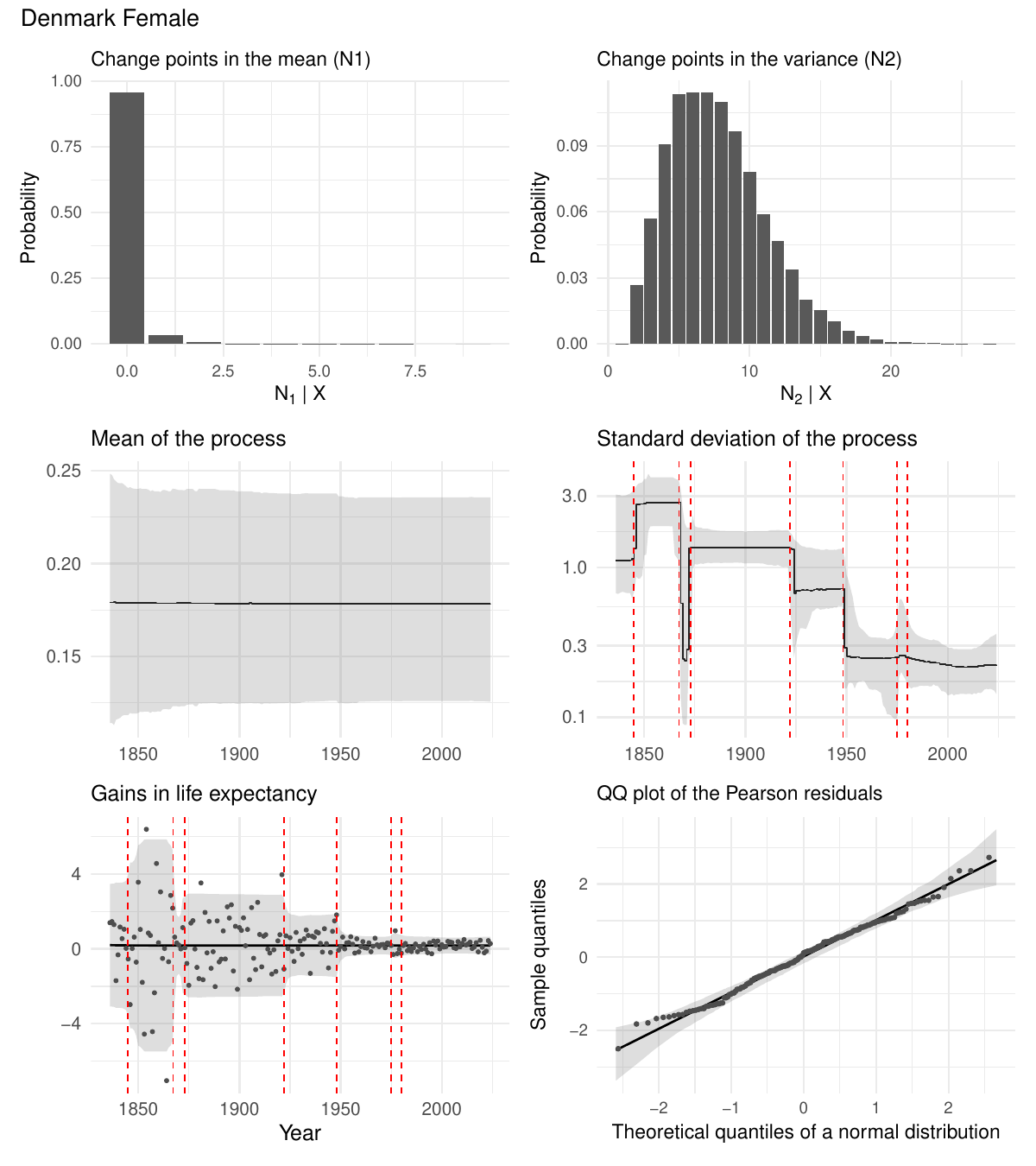}
        \caption{\textbf{Model summaries for annual life expectancy gains in Danish females.} Top panels show posterior probabilities for the number of change points in the process mean ($N_{1}|X$) and variance ($N_{2}|X$). Middle panels display the estimated mean and standard deviation of the process. Solid lines indicate the posterior mode; shaded regions denote 95\% highest posterior density intervals. The bottom-left panel plots observed life expectancy gains (points) against the fitted model, with red dashed vertical lines marking estimated change point years. The bottom-right panel presents a QQ plot of Pearson residuals.}
    \label{fig:denmark_female_0}
\end{figure*}

\begin{figure*}[htbp]
    \centering
    \includegraphics[width=\textwidth]{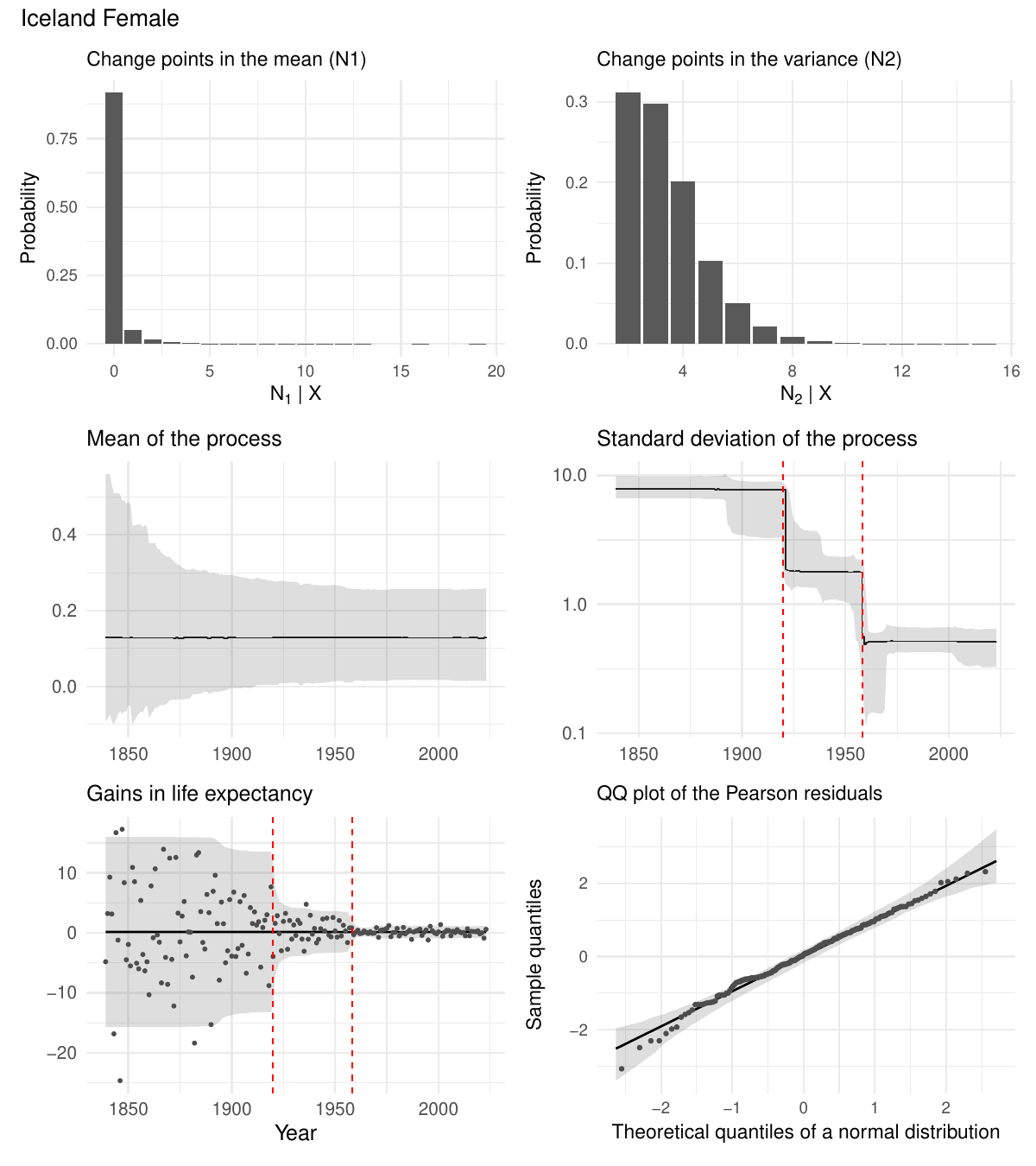}
        \caption{\textbf{Model summaries for annual life expectancy gains in Icelandic females.} Top panels show posterior probabilities for the number of change points in the process mean ($N_{1}|X$) and variance ($N_{2}|X$). Middle panels display the estimated mean and standard deviation of the process. Solid lines indicate the posterior mode; shaded regions denote 95\% highest posterior density intervals. The bottom-left panel plots observed life expectancy gains (points) against the fitted model, with red dashed vertical lines marking estimated change point years. The bottom-right panel presents a QQ plot of Pearson residuals.}
    \label{fig:iceland_female_0}
\end{figure*}

\begin{figure*}[htbp]
    \centering
    \includegraphics[width=\textwidth]{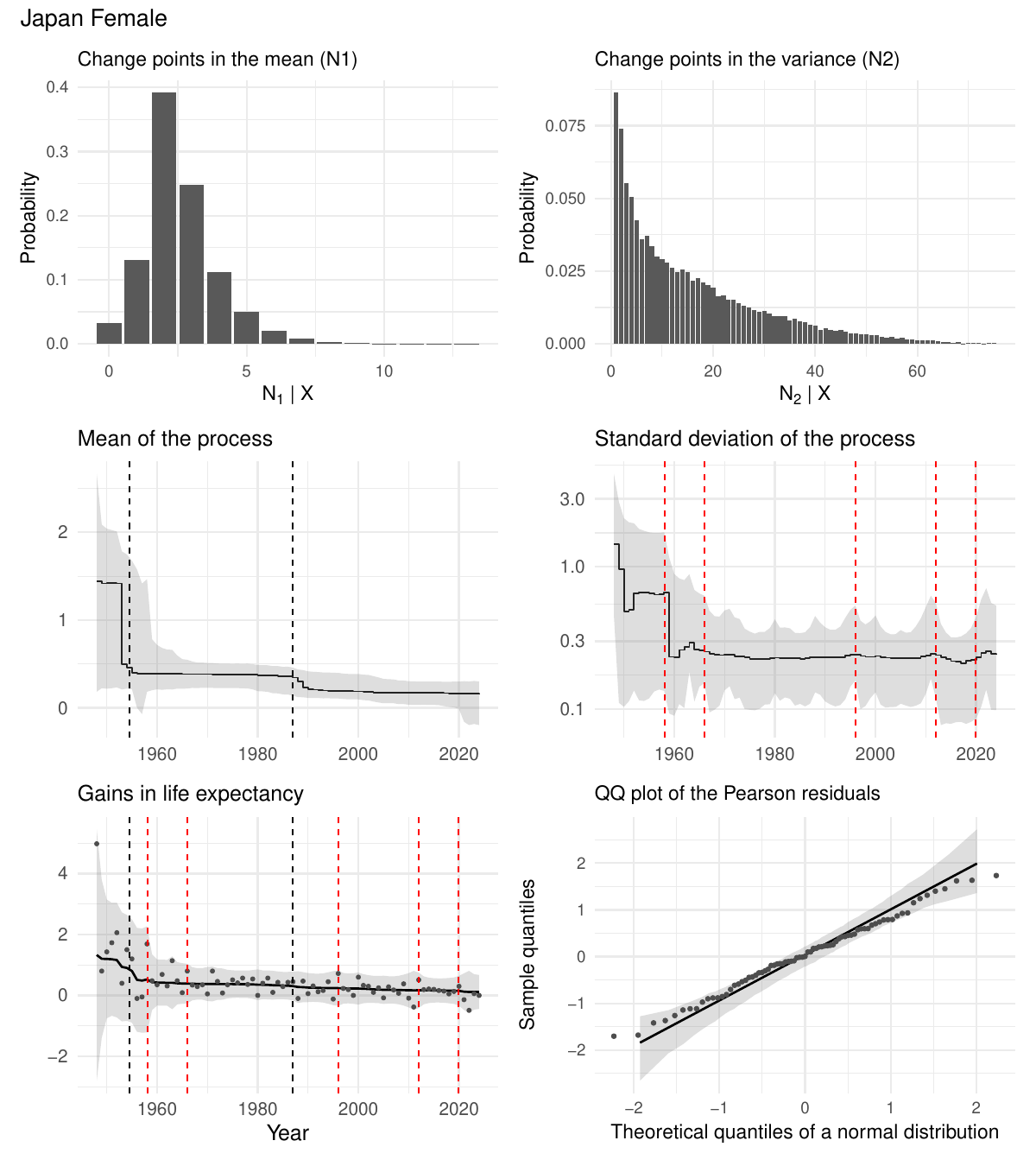}
        \caption{\textbf{Model summaries for annual life expectancy gains in Japanese females.} Top panels show posterior probabilities for the number of change points in the process mean ($N_{1}|X$) and variance ($N_{2}|X$). Middle panels display the estimated mean and standard deviation of the process. Solid lines indicate the posterior mode; shaded regions denote 95\% highest posterior density intervals. The bottom-left panel plots observed life expectancy gains (points) against the fitted model, with red dashed vertical lines marking estimated change point years. The bottom-right panel presents a QQ plot of Pearson residuals.}
    \label{fig:japan_female_0}
\end{figure*}

\begin{figure*}[htbp]
    \centering
    \includegraphics[width=\textwidth]{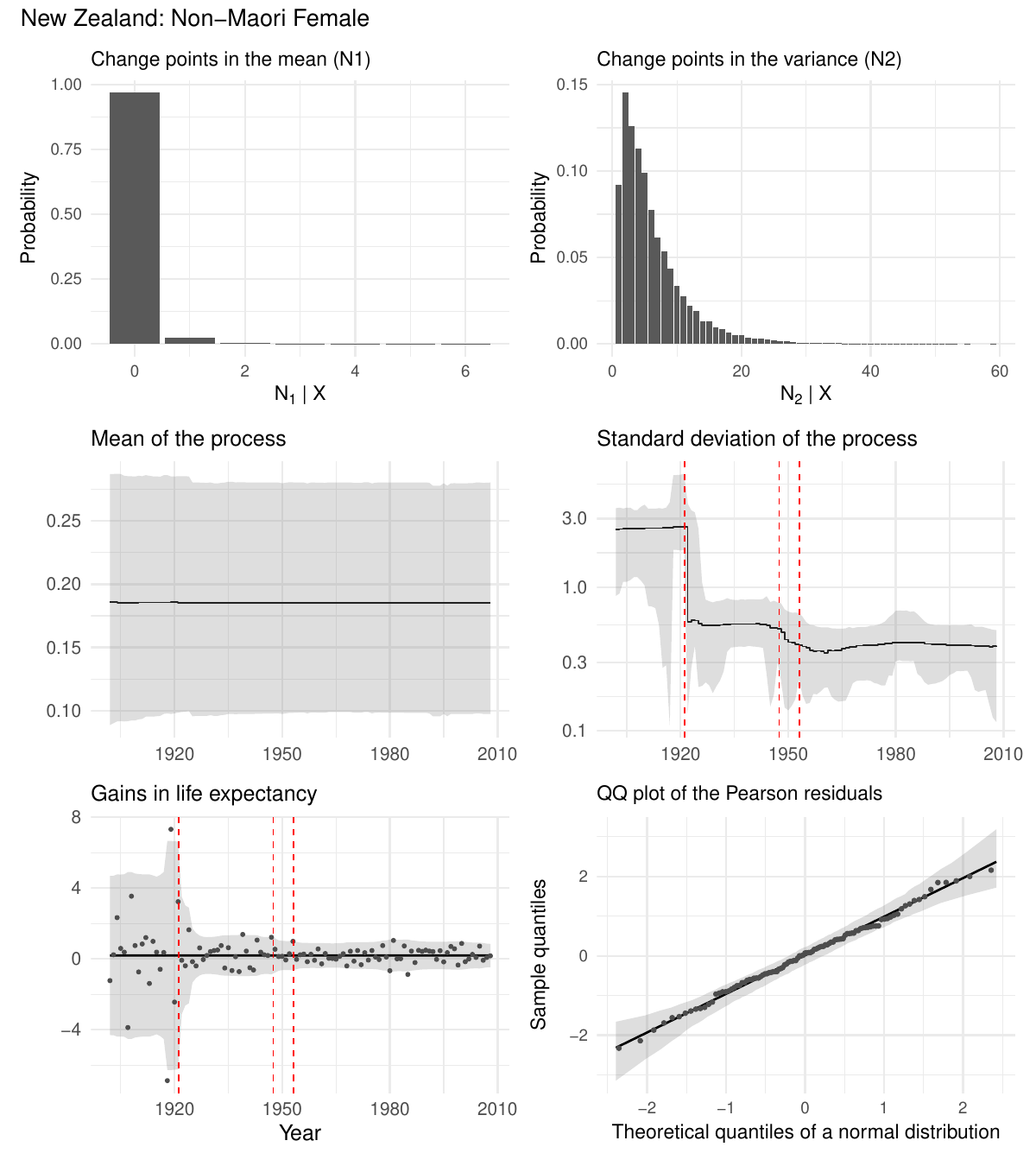}
        \caption{\textbf{Model summaries for annual life expectancy gains in New Zealand females.} Top panels show posterior probabilities for the number of change points in the process mean ($N_{1}|X$) and variance ($N_{2}|X$). Middle panels display the estimated mean and standard deviation of the process. Solid lines indicate the posterior mode; shaded regions denote 95\% highest posterior density intervals. The bottom-left panel plots observed life expectancy gains (points) against the fitted model, with red dashed vertical lines marking estimated change point years. The bottom-right panel presents a QQ plot of Pearson residuals.}
    \label{fig:new_zealand_non_maori_female_0}
\end{figure*}

\begin{figure*}[htbp]
    \centering
    \includegraphics[width=\textwidth]{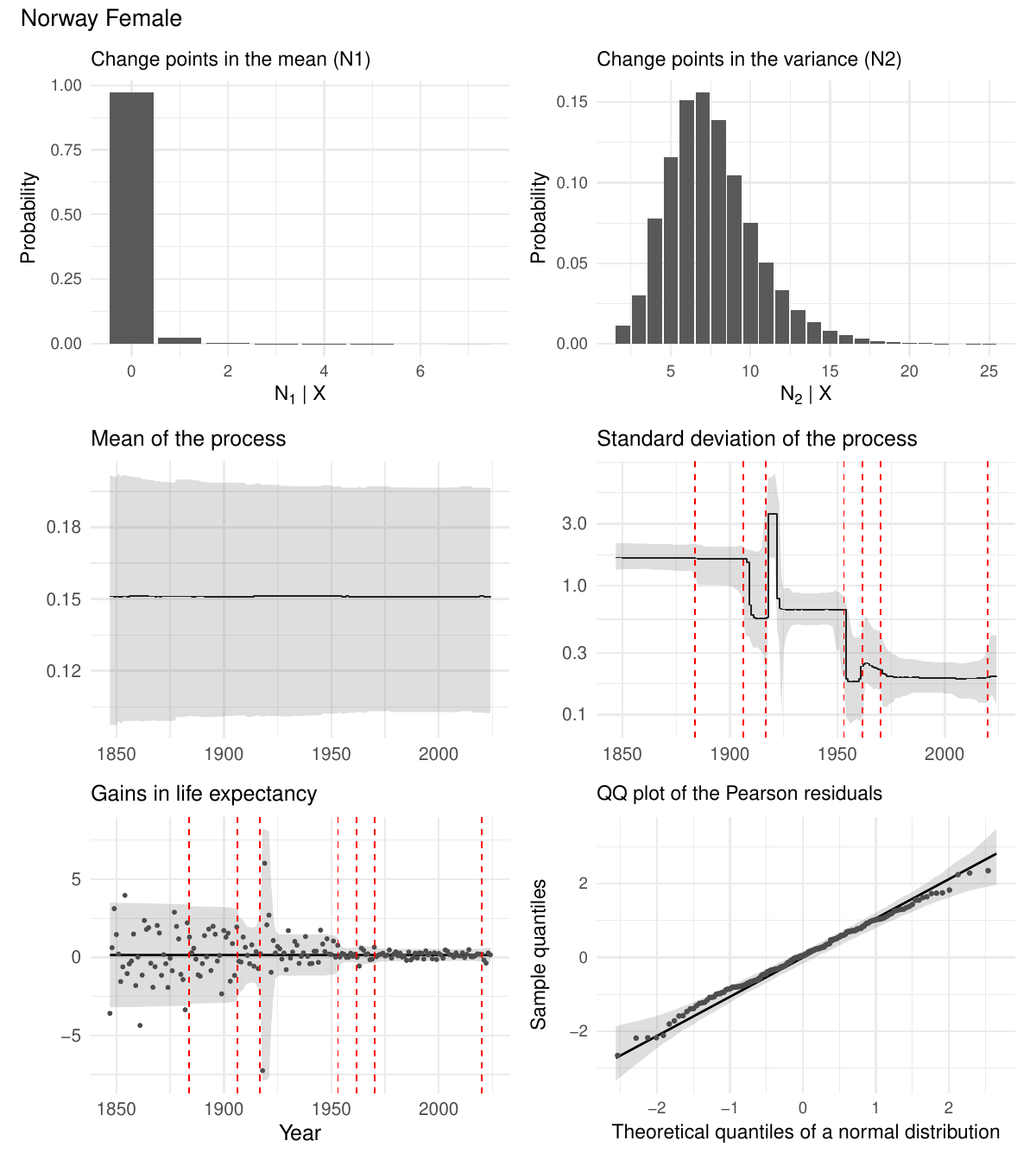}
        \caption{\textbf{Model summaries for annual life expectancy gains in Norwegian females.} Top panels show posterior probabilities for the number of change points in the process mean ($N_{1}|X$) and variance ($N_{2}|X$). Middle panels display the estimated mean and standard deviation of the process. Solid lines indicate the posterior mode; shaded regions denote 95\% highest posterior density intervals. The bottom-left panel plots observed life expectancy gains (points) against the fitted model, with red dashed vertical lines marking estimated change point years. The bottom-right panel presents a QQ plot of Pearson residuals.}
    \label{fig:norway_female_0}
\end{figure*}

\begin{figure*}[htbp]
    \centering
    \includegraphics[width=\textwidth]{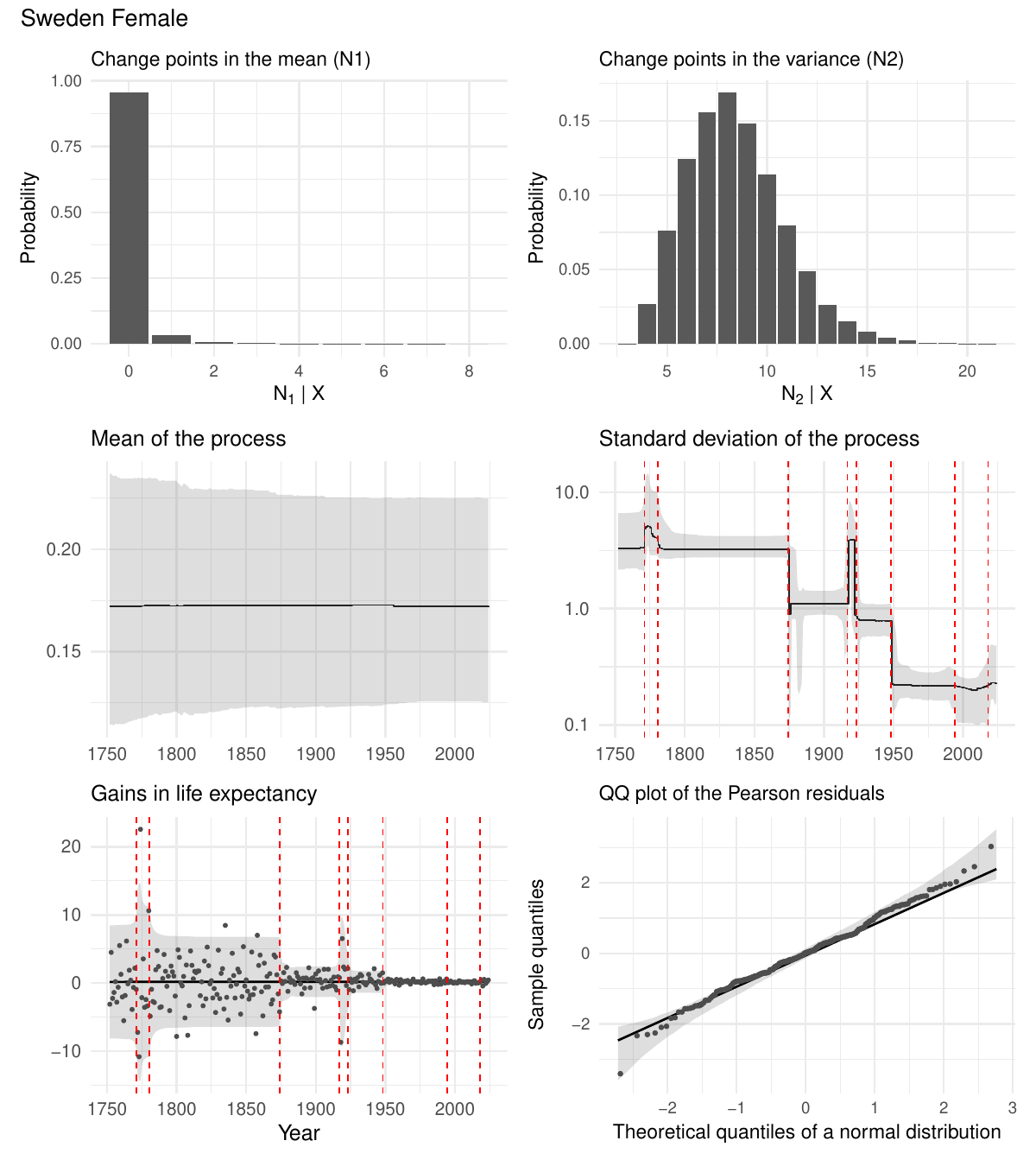}
        \caption{\textbf{Model summaries for annual life expectancy gains in Swedish females.} Top panels show posterior probabilities for the number of change points in the process mean ($N_{1}|X$) and variance ($N_{2}|X$). Middle panels display the estimated mean and standard deviation of the process. Solid lines indicate the posterior mode; shaded regions denote 95\% highest posterior density intervals. The bottom-left panel plots observed life expectancy gains (points) against the fitted model, with red dashed vertical lines marking estimated change point years. The bottom-right panel presents a QQ plot of Pearson residuals.}    \label{fig:sweden_female_0}
\end{figure*}

For each record-holding country, we show the same five diagnostic panels: posterior distributions of the number of change points in the mean ($N_1$) and variance ($N_2$); posterior estimates of $\mu$ and $\sigma$ over time; observed annual gains with the posterior mean and 95\% credible interval; a QQ plot of the Pearson residuals; and observed life expectancy with its OLS trend (Figures \ref{fig:denmark_female_0}--\ref{fig:sweden_female_0}).

\end{appendices}

\end{document}